\documentclass[aps,pra,twocolumn,groupedaddress,superscriptaddress,floatfix,showpacs]{revtex4-1}

\usepackage{graphicx}
\usepackage{amssymb, amsmath}
\usepackage{xcolor}

\usepackage[utf8]{inputenc}

\newcommand{\be}{\begin{equation}}
\newcommand{\ee}{\end{equation}}

\newcommand{\ew}[1]{\left\langle #1\right\rangle}

\newcommand{\dg}{\dagger}
\newcommand{\ket}[1]{\left|#1\right\rangle}

\newcommand{\flip}[2]{\left|#1\right\rangle\left\langle #2\right|}

\begin{document}

\title{Transition from Jaynes-Cummings to Autler-Townes ladder in a quantum dot-microcavity system}

\author{Caspar Hopfmann}
\affiliation{Institute of Solid State Physics, Technische Universität Berlin, D-10623 Berlin, Germany}

\author{Alexander Carmele}
\affiliation{Institut für Theoretische Physik, Nichtlineare Optik und Quantenelektronik, Technische Universität Berlin, Hardenbergstrasse 36, 10623 Berlin, Germany}

\author{Anna Musiał}
\affiliation{Institute of Solid State Physics, Technische Universität Berlin, D-10623 Berlin,
Germany}

\author{Christian Schneider}
\affiliation{Technische Physik, Physikalisches Institut and Wilhelm-Conrad-Röntgen-Resarch Center for Complex Material Systems, Universität Würzburg, D-97074 Würzburg, Germany}

\author{Martin Kamp}
\affiliation{Technische Physik, Physikalisches Institut and Wilhelm-Conrad-Röntgen-Resarch Center for Complex Material Systems, Universität Würzburg, D-97074 Würzburg, Germany}

\author{Sven Höfling}
\affiliation{Technische Physik, Physikalisches Institut and Wilhelm-Conrad-Röntgen-Resarch Center for Complex Material Systems, Universität Würzburg, D-97074 Würzburg, Germany}
\affiliation{SUPA, School of Physics and Astronomy, University of St Andrews, KY16 9SS St. Andrews, United Kingdom}

\author{Andreas Knorr}
\affiliation{Institut für Theoretische Physik, Nichtlineare Optik und Quantenelektronik, Technische Universität Berlin, Hardenbergstrasse 36, 10623 Berlin, Germany}

\author{Stephan Reitzenstein}
\email[]{stephan.reitzenstein@physik.tu-berlin.de}
\affiliation{Institute of Solid State Physics, Technische Universität Berlin, D-10623 Berlin, Germany}


\date{\today}

\begin{abstract}

We study experimentally and theoretically a coherently-driven strongly-coupled quantum dot-microcavity system. Our focus is on physics of the unexplored intermediate excitation regime where the resonant laser field dresses a strongly-coupled single exciton-photon (polariton) system resulting in a ladder of laser-dressed Jaynes-Cummings states. In that case both the coupling of the emitter to the confined light field of the microcavity and to the light field of the external laser are equally important, as proved by observation of injection pulling of the polariton branches by an external laser. This intermediate interaction regime is of particular interest since it connects the purely quantum mechanical Jaynes-Cummings ladder and the semi-classical Autler-Townes ladder. Exploring the driving strength-dependence of the mutually coupled system we establish the maximum in the resonance fluorescence signal to be a robust fingerprint of the intermediate regime and observe signatures indicating the laser-dressed Jaynes-Cummings ladder. In order to address the underlying physics we excite the coupled system via the matter component of fermionic nature undergoing saturation - in contrast to commonly used cavity-mediated excitation. 

\end{abstract}


\maketitle



\section{Introduction}

Advances in realization of quantum technologies and quantum networks rely crucially on the availability of light-matter interfaces, which allow for the initialization, coherent control, read-out and inter-conversion of qubits. Related concepts were first developed and realized in atomic cavity quantum electrodynamics (cQED) \cite{Sch1974, Cir1997, Hoo1998, Nog1999, Rau1999, Mon2002, Bir2005, Fel2006, Moe2007, Kim2008} and later in superconducting circuit QED systems \cite{Wal2004a, Sch2007, Fin2008, Sil2009, Bis2009}. Also, semiconductor-based strongly-interacting light-matter interfaces, which are very appealing in terms of up-scaling and integration, have been demonstrated \cite{Ima1999, Rei2004, Xu2007a, Kas2010, Vol2012, Din2016}. Resonance fluorescence (RF) of strongly interacting systems, consisting of the fundamental cavity mode (FM) of a photonic microcavity and a single quantum dot exciton (QD X), is particularly exciting since it allows for a coherent control of the associated quasi-particle - as shown in a number of recent experiments~\cite{Kas2010, Din2016,Eng2007b, Muel2014, Eng2010, Kim2014, Ota2015, Gre2015, Fis2016}. Interestingly, while the physics of the limiting cases of a) strongly-coupled X-FM systems and the related vacuum Rabi-splitting (VRS) and b) coherently-driven excitons dressed by a strong resonant laser field leading to the Mollow triplet have been studied independently \cite{Rei2004,Yos2004,Ate2009,Jah2012}, the intermediate regime of strong coherent driving of strongly-coupled exciton-cavity system has not been explored so far. In this regime, which is subject of this work, the excitation laser strength $g_\mathrm{L-X}$ becomes comparable to the light-matter coupling strength $g_\mathrm{X-FM}$, and therefore the behavior of the system is qualitatively different from the limiting cases in which one of the couplings dominates the system and the other can be treated as a weak perturbation. As a result, the observation of laser-dressed polaritonic states is expected. This raises important question of how far it is possible to climb the Jaynes-Cummings ladder (so far limited to signatures of up to the 2nd rung in QD-based cQED systems \cite{Eng2007b, Far2008, Fin2008, Rei2012b, Kas2010}) before it becomes dressed by the coherent driving. It is also related to the question if and under which excitation conditions a single-QD laser can be realized \cite{McKee2003, Nod2006, Rei2008b, Nom2009, Str2011}. Furthermore, it is relevant for the recent investigations of the transition from strong coupling to lasing \cite{Gie2016}. In our approach, this interesting prospect could potentially be enabled by highly selective and efficient resonant excitation of the QD exciton. As such, the evolution of the occupation of the coherently-driven strongly-coupled X-FM system and its eigenstates with increasing driving strength is of fundamental interest for the field of cQED. Examples are the discussion about the observability of higher order Jaynes-Cummings rungs and their transformation into the laser-dressed Jaynes-Cummings ladder under coherent excitation as well as single-QD lasing.

\begin{figure}
	\includegraphics[width=1.0\linewidth]{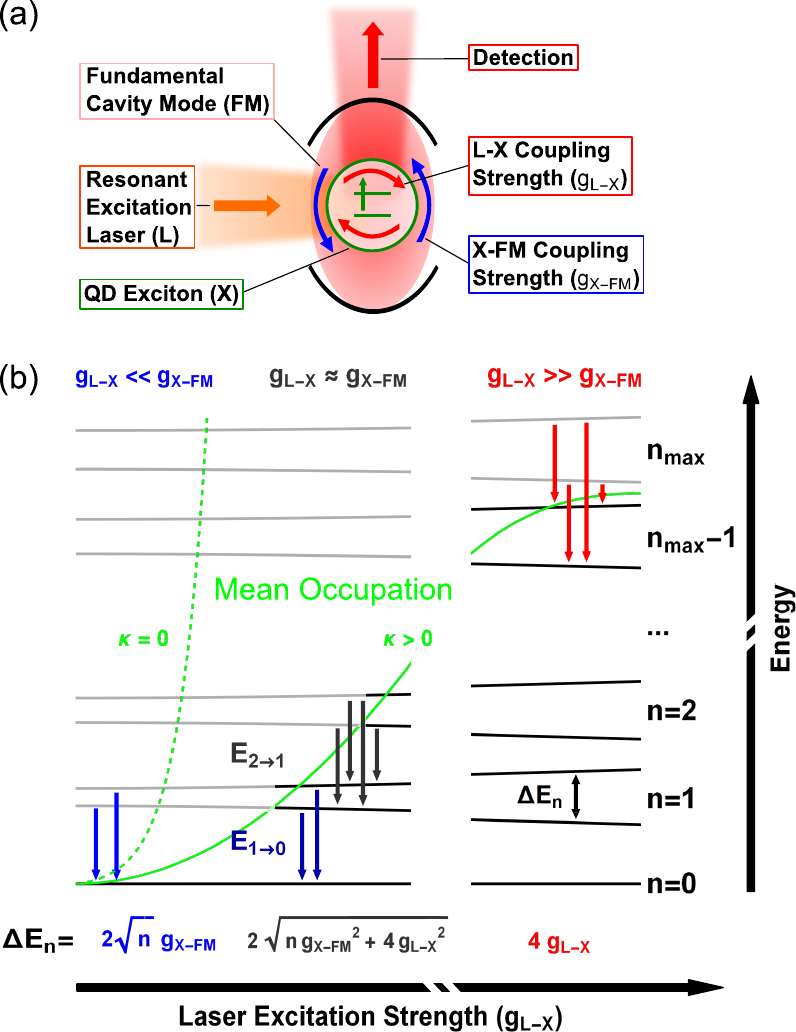}
	
	\caption{\label{fig:Schematic_Config} (a) Scheme of a quantum dot (QD)-microcavity system resonantly excited in lateral direction. The fundamental microcavity mode (FM) is oriented orthogonally with respect to the excitation laser field (L), as a consequence the QD exciton X-FM system is resonantly excited via the X. The coupling strengths between L and X and between X and FM are denoted with $g_\mathrm{L-X}$ and $g_\mathrm{X-FM}$, respectively. (b) Level scheme of the laser dressed X-FM system neglecting dephasing as a function of $g_\mathrm{L-X}$. The system mean occupation is indicated as a guide to the eye in green. The possible transitions between states differing in excitation manifold $n$ are depicted by arrows (e.g., from manifold $n$ to $n-1$ are denoted by $E_{n \to n-1}$). The splittings of the manifolds $\Delta E_n(g_\mathrm{L-X})$ in different regimes are given by formulas at the bottom. The limiting cases of low and high excitation $g_\mathrm{L-X}$ simplify to pure Jaynes-Cummings (blue) and Autler-Townes (red) ladder transitions, respectively. The green curves indicate climbing up the Jaynes-Cummings ladder for negligible photonic losses (dashed curve, $\kappa = 0$), and the transition from the anharmonic Jaynes-Cummings ladder to the harmonic Autler-Townes ladder when significant losses are present (solid curve, $\kappa > 0$).}
\end{figure}

In this work we address experimentally and theoretically a coherently-driven strongly-coupled cQED system (Fig. \ref{fig:Schematic_Config}(a)) and focus on the regime of mutual strong coupling between three oscillators: the laser light field L, the quantum dot exciton X, and the (fundamental) cavity mode of a microresonator FM beyond the description of the limiting cases when one of the couplings dominates. Thereby, we investigate the influence of the interplay between the coupling strengths $g_\mathrm{L-X}$ and $g_\mathrm{X-FM}$ on the optical response of the system under resonant driving. We define the conditions to observe the intermediate regime. Namely, we identify driving the system through the matter state as well as the ratio between X-C coupling strength and cavity losses as crucial factors. Interestingly, our results indicate that dressing of the polariton is not possible if the system is excited through the cavity mode due to its unlimited occupation (bosonic reservoir), But that it is a unique feature of the direct driving of the X undergoing saturation. We further examine in detail the differences and consequences of the nature of the state through which the system is excited (bosonic C versus fermionic X). Depending through which state the system is driven this leads to a different system evolution with increasing excitation strength and fundamentally different physical system in the strong driving regime. 

In our excitation scheme the coupled X-FM polariton is excited by a resonant laser which is tuned to the energy of the bare (uncoupled) X transition. This is a distinctive feature of our work in comparison to commonly used cavity-mediated excitation \cite{Din2016, Gre2015, Car2015}. We describe and exploit the significant difference in the nature and lifetime of the state through which the system is pumped, i.e. a difference in the range of two orders of magnitude between X lifetime of ($0.35 - 1$) $ns$ and cavity photon lifetime in the range of ($5 - 10$) $ps$. Interestingly, the investigated system exhibits drastically different character depending on the driving amplitude $g_\mathrm{L-X}$: In the case of weak driving ($g_\mathrm{L-X} << g_\mathrm{X-FM}$) the laser is only probing the X-FM polaritons which form if $g_\mathrm{X-FM}$ is large enough to overcome the losses \cite{Rei2004}. In this regime the vacuum Rabi dublet is observed, with an upper and lower polariton (UP, LP respectively) (cf. left in Fig. \ref{fig:Schematic_Config}(b)). On the other hand in the limit of strong coherent driving ($g_\mathrm{L-X} >> g_\mathrm{X-FM}$) of X the resonant laser dresses the X state resulting in the Autler-Townes splitting of both ground and excited state proportional to the driving strength - the Rabi splitting of $4 \, g_\mathrm{L-X}$ (Fig. \ref{fig:Schematic_Config}(b) right). This results in the characteristic three-peak Mollow triplet structure in the spectrum \cite{Mol1969, Ate2009}. The unexplored transitory regime (Fig. \ref{fig:Schematic_Config}(b) center) in which $g_\mathrm{L-X} \approx g_\mathrm{X-FM}$ is the principal topic of this study.

The paper is organized as follows: Information regarding the employed QD-microcavity structure and experimental setup, as well as basic characterization of the X-FM system is given in Sec. \ref{Sec: Methods}. Sec. \ref{Sec: Theo} introduces the theoretical model and presents calculated spectral response of the mutually coupled system as a function of excitation power. Additionally, the differences of X- and FM-driving schemes are evaluated. Sec. \ref{Sec: Exp} presents a discussion of experimental results on the RF of coherently-driven strongly-coupled X-FM polariton as well as excitation power-dependent measurements. Furthermore, the experimental results are contrasted to theory. A summary of our findings is provided in Sec. \ref{Sec: Conclusion}.


\section{Methods}
\label{Sec: Methods}

As model structures for experimental realization of a coherently-driven strongly-coupled cQED system we use high-quality (average Q-factor of $13 000$) low mode-volume ($0.43$ $\mu m^3$) micropillar cavities \cite{Rei2007}. These structures, based on laterally-extended self-assembled InGaAs QDs with high oscillator strength in the range of $20$-$50$ \cite{Rei2004, Rei2009a}, enable the realization of strong coupling between single Xs and single FMs with exceptional $50 \%$ yield \cite{Hop2015} and $g_\mathrm{X-FM}$ up to $65$ $\mu eV$.

\begin{figure}
	\includegraphics[width=0.9\linewidth]{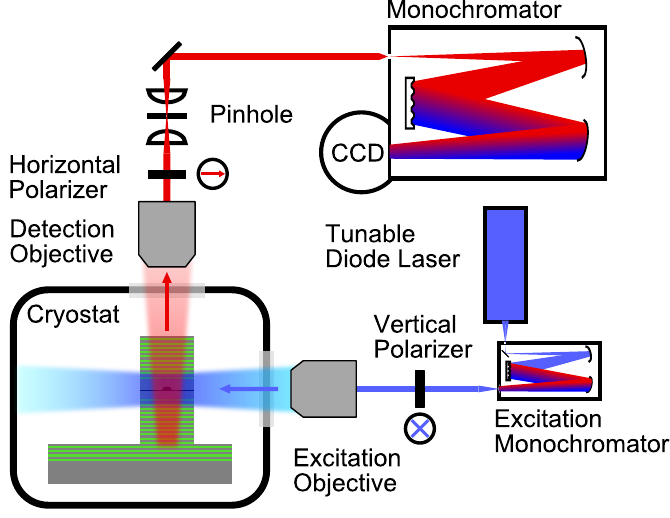}
	\caption{\label{fig:Suppl_Setup} Scheme of the microphotoluminescence experimental setup enabling in-plane excitation of the QD-micropillars and providing efficient laser stray-light suppression for RF experiments (elements are not to scale).}
\end{figure}

To experimentally realize the scenario of the X-driven laser-dressed polariton system and to enable the observation of the limiting excitation regimes, a $90^{\circ}$ lateral excitation and vertical detection setup is used. This enables resonant access to the X which is not modulated by the wavelength-dependent reflectivity of the microcavity mirrors. Ideally, the optical field of the laser does not interact with the FM due to $90^{\circ}$ orientation of the laser propagation direction and the optical field of the FM providing enhanced suppression of scattered laser light crucial for RF experiments \cite{Ota2015, Ate2009a}. Since micropillar cavities feature solid state material interfaces - such as the lateral cavity boundary and DBR layers - on which the impending excitation beam may scatter, residual stray-light is detected even in a $90^\circ$ excitation/detection scheme. The experimental setup is shown schematically in Fig. \ref{fig:Suppl_Setup}. For resonant excitation a narrow linewidth ($< 100$ $kHz$) tunable diode laser and for above-band excitation a $532$ $nm$ frequency doubled neodymium-doped yttrium aluminum garnet (Nd:YAG) solid state laser (not shown) are employed. The resonant laser light is guided through a monochromator (bandwidth of about $0.015$ $nm$ or $21$ $\mu eV$ at $930$ $nm$) in order to suppress the LED-like background emission typical for tunable diode lasers. The resonant excitation is polarized orthogonally with respect to the QDs growth as well as the micropillar cavity axis. The detection signal is filtered by a linear polarizer in the orthogonal direction of the excitation to enhance the stray-light suppression, further suppression is achieved by usage of a pinhole as a spatial filter to limit detected signal to radiation passing through the top facet of the micropillar. The excitation and detection objectives can be adjusted independently and feature numerical apertures of $0.4$ and $0.65$, respectively, providing spatial resolution in the range of $2$ $\mu m$. A spectrometer, consisting of a $0.75$ $m$ focal length monochromator and nitrogen-cooled Si charge-coupled device, is used to analyze the detected light with the spectral resolution of about $25$ $\mu eV$ at $930$ $nm$. The sample is mounted in helium-flow cryostat and cooled down to temperatures in the range of $5$ $K$ to $60$ $K$.\\

\begin{figure}
	\includegraphics[width=0.8\linewidth]{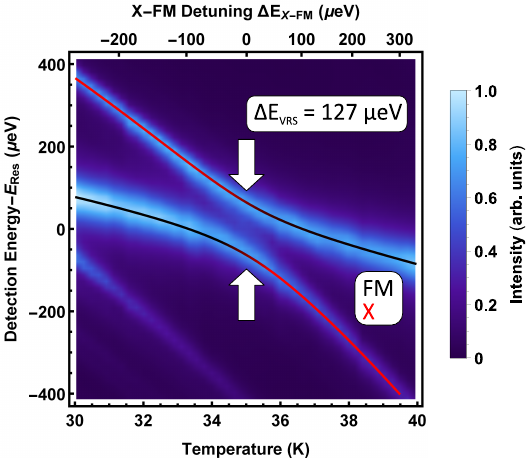}
	\caption{\label{fig:Suppl_TempScan} Temperature-dependent photoluminescence spectra under low above-band excitation at $532$ $nm$ with $15$ $\mu W$ power $P$ measured outside the cryostat, together with fitted quantum dot exciton (X, red solid line) and fundamental cavity mode (FM, black solid line) emission energies. The color change of the lines around the resonance energy $E_\mathrm{res}$ corresponds to exchange of the X and FM characteristics. We determine a vacuum Rabi splitting of 127 $\mu$eV on resonance.}
\end{figure}

The basic parameters of the strongly-coupled X-FM system are determined from above-band excitation photoluminescence measured as a function of X-FM detuning (Fig. \ref{fig:Suppl_TempScan}). The temperature is utilized to tune the X through the FM using their different temperature dispersions \cite{Rei2004}. Depicted curve shows pronounced anti-crossing typical for strongly-coupled systems and was fitted with a global (2D) fit according to \cite{Lau2008}. The values of the X-FM coupling strength  $g_\mathrm{X-FM} = 65$ $\mu eV$, resonance temperature $T_\mathrm{res} = 35.0$ $K$, FM full width at half maximum (FWHM) $\kappa = 110$ $\mu eV$ and FWHM of the QD exciton $\gamma_{X}^\mathrm{532nm} = 52$ $\mu eV$ can be determined using this model. The FWHM of the QD is decreased to about $\gamma_X = 15$ $\mu eV$ under resonant excitation due to reduced dephasing \cite{Ben2005c, Ate2012}. The characteristic feature of a strongly-coupled system is the vacuum Rabi splitting (VRS) determined on resonance to be equal to $\Delta E_\mathrm{VRS} = 127$ $\mu eV \lesssim 2 g_\mathrm{X-FM}$ thereby accounting for photonic losses, but no additional linewidth broadening mechanisms - such as spectral jitter \cite{Lau2008, Mus2015, Rei2004}. The FM - which is ideally twofold degenerate - exhibits a mode splitting $\delta$FM of about $16$ $\mu eV$ into linearly cross-polarized mode components FM1 and FM2, which can be attributed to slight asymmetry of the micropillar cross-section \cite{Rei2010c, Ate2007}.


\section{Theory}
\label{Sec: Theo}

The strongly-coupled X-FM system under resonant excitation is modelled employing a Hamiltonian written in the dipole approximation and rotating frame of the driving laser:

\begin{align}
	H =& H_0+H_\mathrm{L-X}+H_\mathrm{X-FM} \label{eq:H}\\
	H_0 =& \hbar \, \sigma_{ee}  \, \Delta E_\mathrm{L-X} + \hbar \, c^\dagger c \, \Delta E_\mathrm{L-FM} \label{eq:H_0}\\
	H_\mathrm{L-X} =&  g_\mathrm{L-X} \, \mathcal{N}(\gamma_\mathrm{at}, \Delta E_\mathrm{L-X}) \left( \sigma_{ge} + \sigma_{eg} \right) \label{eq:H_L-X} \\
	H_\mathrm{X-FM} =& g_\mathrm{X-FM} \left( c^\dagger \sigma_{ge} + \sigma_{eg} c \right). \label{eq:H_X-FM} 
\end{align}

The X and FM energies relative to the laser energy $E_L$ are denoted with $\Delta E_\mathrm{L-X}$ and $\Delta E_\mathrm{L-FM}$, respectively. The transition and occupation operators between excited $\ket{e}$ and ground $\ket{g}$ state and of X are expressed with $\flip{i}{j}=\sigma_{ij}$. The system is driven through X by a coherent laser field of strength $g_\mathrm{L-X}$ which is related to the excitation power $P$ by $g_\mathrm{L-X} \propto \sqrt{P}$. $\mathcal{N}(\gamma_\mathrm{at}, \Delta E_\mathrm{L-X})$ is an envelope function of a normal distribution of a FWHM of $\gamma_\mathrm{at}$ which models the excitation laser attenuation as a function of the $\Delta E_\mathrm{L-X}$ detuning. By applying this function we phenomenologically describe the experimental observation that the polariton states are not efficiently pumped by the resonant laser (see Fig. \ref{fig:RF_LScan}(a)). Note that for resonant power dependent studies of the transitory regime (i.e. Figs. 1, 4, 5, 7 and 8) the attenuation equals to $\mathcal{N} = 1$ and is therefore irrelevant. To model the experimental results we calculate the spectra from two-time correlations of the cavity operators obtained via a master equation in the standard Lindblad form, by taking cavity losses, decoherence and radiative decay into account in the dissipative part of the Hamiltonian $H$. All theoretical data is obtained via a numerical evaluation of the master equation up to very high-orders to include multi-photon scattering events, which are crucial for the description of a coherently-driven cQED setup. In our case none of the interactions $H_\mathrm{L-X}$ or $H_\mathrm{X-FM}$ can be treated perturbatively. Details are given in appendix \ref{App: Theo Spectra}.\\


\begin{figure}
	\includegraphics[width=1.0\linewidth]{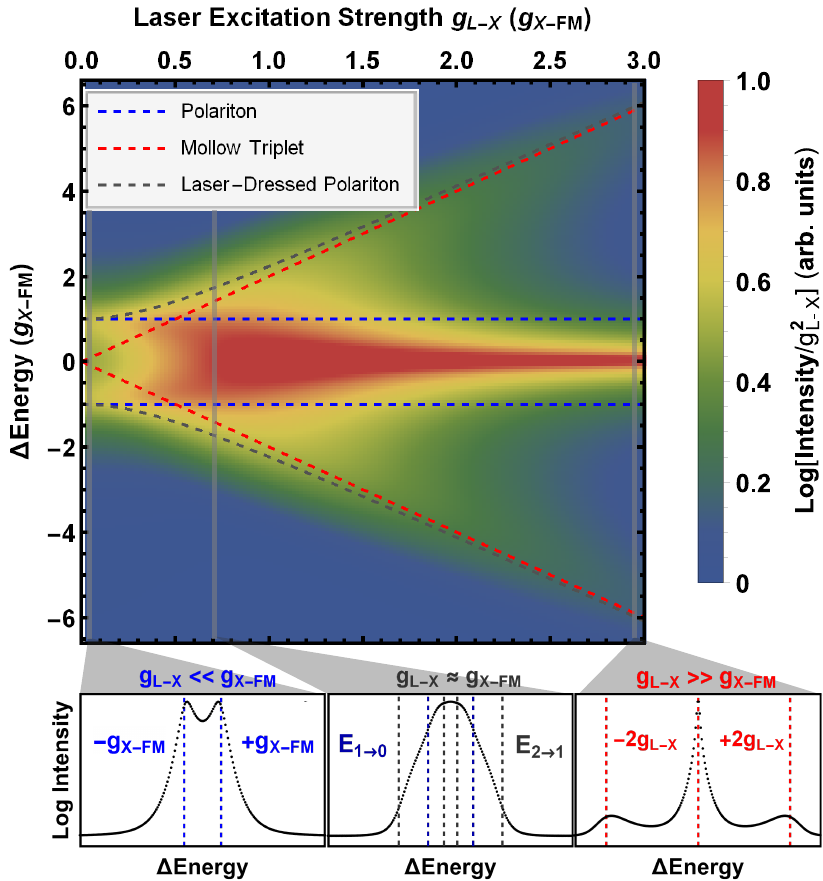}
	\caption{\label{fig:Theo_PScan} Driving strength $g_\mathrm{L-X}$-dependent theoretical incoherent emission spectra under resonant excitation normalized by $g_\mathrm{L-X}^2$. QD exciton (X), fundamental mode (FM) and laser (L) are in resonance with each other. The eigenstates (neglecting dephasing) of X-FM polariton $= \pm g_\mathrm{X-FM}$, L-X Mollow triplet $= \pm 2 \, g_\mathrm{L-X}$ and laser-dressed polariton $= \pm \sqrt{g_\mathrm{X-FM}^2+\, 4 g_\mathrm{L-X}^2}$ (derived as shown in appendix \ref{App: Theo Ana Eigen}) are drawn as a guide to the eye. Three exemplary spectra (in three different excitation regimes) are shown below the intensity map with relevant transitions (corresponding to level scheme in Fig. \ref{fig:Schematic_Config}(b)) indicated by dashed lines.}
\end{figure}

In order to characterize the transition between the limiting anharmonic Jaynes-Cummings and the harmonic Autler-Townes ladders of the low and high coherent driving regime (cf. \ref{fig:Schematic_Config}(b)), the calculated FM spectrum is investigated as a function of the driving strength $g_\mathrm{L-X}$. When increasing $g_\mathrm{L-X}$ we aim at accessing the intermediate regime (center region in Fig. \ref{fig:Schematic_Config}(b)) in which $g_\mathrm{L-X} \approx g_\mathrm{X-FM}$ and the laser can no longer be treated as weak perturbation of the strongly-coupled X-FM system. In the presented investigation all oscillators are on resonance, i.e. $\Delta E_\mathrm{L-X} = \Delta E_\mathrm{X-FM} = 0$. For generality all energies in the system are expressed relative to $g_\mathrm{X-FM}$. The calculated $g_\mathrm{L-X}^2$-normalized incoherent cavity spectra, using experimentally-determined system parameters, are depicted as a function of $g_\mathrm{L-X}$ in Fig. \ref{fig:Theo_PScan}. The normalization by $g_\mathrm{L-X}^2$ is introduced to keep experimental (cf. Fig. \ref{fig:Exp_PScan}) and theoretical results comparable, which is necessary since in experiment there is a finite background from excitation stray-light scaling with $P \propto g_\mathrm{L-X}^2$. At low excitation below $0.1$ $g_\mathrm{X-FM}$ the coupling between X and FM is only weakly perturbed by the resonant laser field providing system occupation, but not changing the eigenstates of the system and, therefore, a standard VRS is observed. In the high excitation regime above $1.0$ $g_\mathrm{X-FM}$ the system is dominated by the laser dressing and a Mollow triplet with a splitting between its sidebands of $ = \pm 2 \, g_\mathrm{L-X}$ emerges. In the intermediate regime between $0.1$ and $1.0$ $g_\mathrm{X-FM}$ when $g_\mathrm{L-X} \approx \frac{g_\mathrm{X-FM}}{2}$ the system response resembles that of three (equally) strongly-coupled oscillators and the X-FM polariton is dressed by the laser thereby forming a quasi particle consisting of two photons and one exciton. For higher cavity occupations we are dealing with a ladder of dressed states with excitation strength-dependent (tunable) splittings and anharmonicity inherited from the Jaynes-Cummings ladder which is fundamentally different from the polariton and laser-dressed QD X limiting cases.

Three exemplary spectra (vertical cross sections) in the different excitation regimes are shown below the intensity map, relevant transitions (e.g. for manifold $n$: $E_{n \to n-1}$) corresponding to the simplified level scheme of Fig. \ref{fig:Schematic_Config}(b) are indicated. The level scheme is able to explain the observed spectra qualitatively. In ideal systems without dissipation, the VRS equals to $2 \, g_\mathrm{X-FM}$ and increasing incoherent excitation strength leads to the formation of higher excitation manifolds $n$ with splittings scaling with $\sqrt{n}$ - the Jaynes-Cummings ladder \cite{Jay1963}. In experimental cQED systems climbing the Jaynes-Cummings ladder has been so far hindered by dephasing \cite{Mue2009, Bis2009, Sea2012}. Interestingly, even in an ideal system under coherent driving it is not possible to climb the Jaynes-Cummings ladder to arbitrarily high states because treating the laser as a weak perturbation is not valid anymore and as a result it influences the system's eigenstates beyond the Jaynes-Cummings model. Below we define the conditions (both experimental and regarding system parameters) for which higher order rungs of Jaynes-Cummings ladder cannot be observed, because instead of climbing up the ladder, its states are dressed and further increase in the excitation strength leads to climbing the ladder of dressed polariton states. Since the spectral widths associated with X ($\gamma_X$) and FM ($\kappa$) are about $0.24$ $g_\mathrm{X-FM}$ and $1.68$ $g_\mathrm{X-FM}$, respectively, it is not possible to spectrally resolve transitions of the dressed polariton states. To reduce overlap between resonances and spectrally resolve the individual transitions cavity losses and X dephasing have to be reduced in order to fulfill $\sqrt{\kappa^2 + \gamma_X^2} << g_\mathrm{X-FM}$, which in our system implies $Q > 250000$ and $\gamma_X < 5$ $\mu eV$. However, reducing $\kappa$ has significant consequences for the X-driven cQED system: $\kappa$ determines at which $g_\mathrm{L-X}$ the transition to the Mollow triplet takes place. When it comes to domination of the laser over the system eigenstates and thus the optical response, the important figure of merit is the ratio $g_\mathrm{X-FM} / \kappa$. As long as higher polariton rungs can be efficiently pumped and populated due to the exciton driving, a Mollow triplet cannot be formed. Only, when $\kappa$ limits the excitation transfer, the coherent driving exceeds the intrinsic time scale of the cQED system and starts to create a laser-dressed state of the quantum dot. In other words, for lower cavity losses $\kappa$ the transitory regime is shifted to higher excitation powers, limiting its observability in both experiment and theory. This assertion and established criteria for realization of the transition to the laser-driven QD system changes view on relatively high cavity losses in cQED systems. The usually unwelcome losses are transformed into a desired attribute which denotes an indispensable condition for observation of the intermediate dressed-polariton regime and the Mollow triplet. As a (counterintuitive) consequence high losses are crucial for formation of a highly coherent laser dressed polariton state. Not only the optical response of the system, but the system itself is very sensitive to the parameters of each oscillator and therefore behaves almost chaotically. Namely even slight changes may result in different regime and in that case fundamentally different physical system and its evolution with the driving strength. This is what up till now hindered the consistent and unified description of different regimes and transition between them. We would like to point out that the saturation of the mean photon number in the exciton-driven system can be analytically estimated to $0.5 \, g_\mathrm{X-FM}^2/(\kappa^2 + g_\mathrm{X-FM}^2)$ which evaluates in our case to about $0.4 < 1$. The latter implies that the system does not reach the regime of (single-QD) lasing even under coherent driving \cite{Bjoe1994}. The eigenstates of the coupled three oscillators in the transitory regime differ significantly from the eigenstates of the coupled X-FM and laser driven-X systems of the low and high excitation limit, respectively. In our case the deviation is strongest at $\frac{g_\mathrm{X-FM}}{2}$, defining a condition for presence of the transitory regime for a given $\kappa$.

\begin{figure}
	\includegraphics[width=0.7\linewidth]{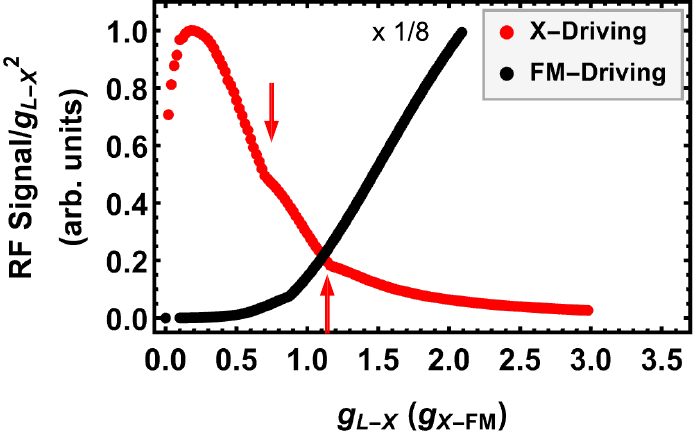}
	\caption{\label{fig:Suppl_ExComp} Comparison of theoretical combined coherent and incoherent resonance fluorescence (RF) intensity integrated over detuning range of $\pm 0.1 g_\mathrm{X-FM}$ as a function of excitation strength $g_\mathrm{L-X}$ for direct excitation of the quantum dot exciton (X, red) and the fundamental mode (FM, black), respectively. X-driving RF intensity curve slope changes - corresponding to the occupation of higher order rungs - are indicated by arrows.}
\end{figure}

In the aforementioned discussion the transition between the Jaynes-Cummings and Autler-Townes ladders is only relevant for X-driven systems, this is illustrated in the following. The response of the strongly-coupled X-FM system to coherent driving depends on whether the resonant laser pumps the system via the X or the FM. This has a very important implication, which has not been fully explored so far, mainly for the excitation efficiency and behavior of the system in the limit of strong driving. As discussed in Sec. \ref{Sec: Methods}, the experimental configuration determines the excitation scheme the system is subjected to. In the case of QD-micropillar cavities excitation mediated via X corresponds to lateral excitation - as realized experimentally in this study. The FM-mediated experiments on the other hand were realized, e.g., in Ref. \cite{Gre2015}. In order to compare the two excitation channels we choose as a figure of merit the RF signal, which is defined as the total spectral intensity normalized by excitation power (RF response containing both the incoherent and coherent part) integrated over the detuning range of $\pm 0.1 g_\mathrm{X-FM}$ around the laser energy. Extracted RF signals of calculated spectra for X- and FM-driving as a function of $g_\mathrm{X-FM}$ are presented in Fig. \ref{fig:Suppl_ExComp}. For the X-driven system the RF signal (red) shows a maximum at an excitation strength $g_\mathrm{L-X}$ of about $0.18$ $g_\mathrm{X-FM}$. The theoretical analysis of the occupation of laser-dressed polariton states indicates that this RF signal maximum corresponds to the population of the first excitation manifold of the mutually coupled system. It can only be observed in the X-driven configuration (see below) and therefore represents a fingerprint of the transitory regime and the formation of the laser-dressed polariton states. Interestingly, the $g_\mathrm{L-X}$ value of the maximum is an inherent feature of the strongly-coupled laser-X-FM system, which is robust against specific system parameters. For both higher and lower $g_\mathrm{L-X}$ the X-driven RF signal drops down to $0$. Towards large $g_\mathrm{L-X}$ two distinct changes in the slope (red arrows) can be observed. The changes can be traced back to photon probabilities and correlation functions (not shown here) exhibiting (only under X-driving) maxima at excitation strengths $g_\mathrm{L-X}$ corresponding to occupation of rungs of the dressed Jaynes-Cummings ladder - $\sqrt{n} \, g_\mathrm{X-FM}$ (neglecting dephasing). The spectral contributions of the higher rungs decrease in amplitude with each occupied manifold $n$. In the calculations up to the 6th manifold has to be included in order to achieve convergence. Calculating the RF signal contributions of each individual  excitation manifold separately is non-trivial and beyond the scope of this work. The FM-driven system RF signal (black) shows a drastically different behavior - it is monotonically increasing ($\propto \overset{n}{\sum }g_{L-X}^{2 n}$) and no maxima can be observed. The $g_\mathrm{L-X}$ range for the FM-driving is limited to values below $2 g_\mathrm{X-FM}$ because the number of higher excitation manifolds that need to be included diverges quickly and so does the calculation time, e.g., at $2 g_\mathrm{X-FM}$ already more than $100$ manifolds need to be included. The principal difference between X- and FM-driving is clearly visible even in this limited range and can be explained in terms of the different character of the state through which the system is pumped, i.e. X and FM - fermionic and bosonic, respectively. As a consequence in X-driving the occupation (and therefore the maximum occupied manifold) is limited by the X lifetime-governed saturation of the electronic state together with the cavity loss rate. While FM-driving the occupation is only limited by the cavity loss rate and therefore diverges with increasing excitation strength.


\section{Experiments}
\label{Sec: Exp}

\begin{figure}
	\includegraphics[width=0.9\linewidth]{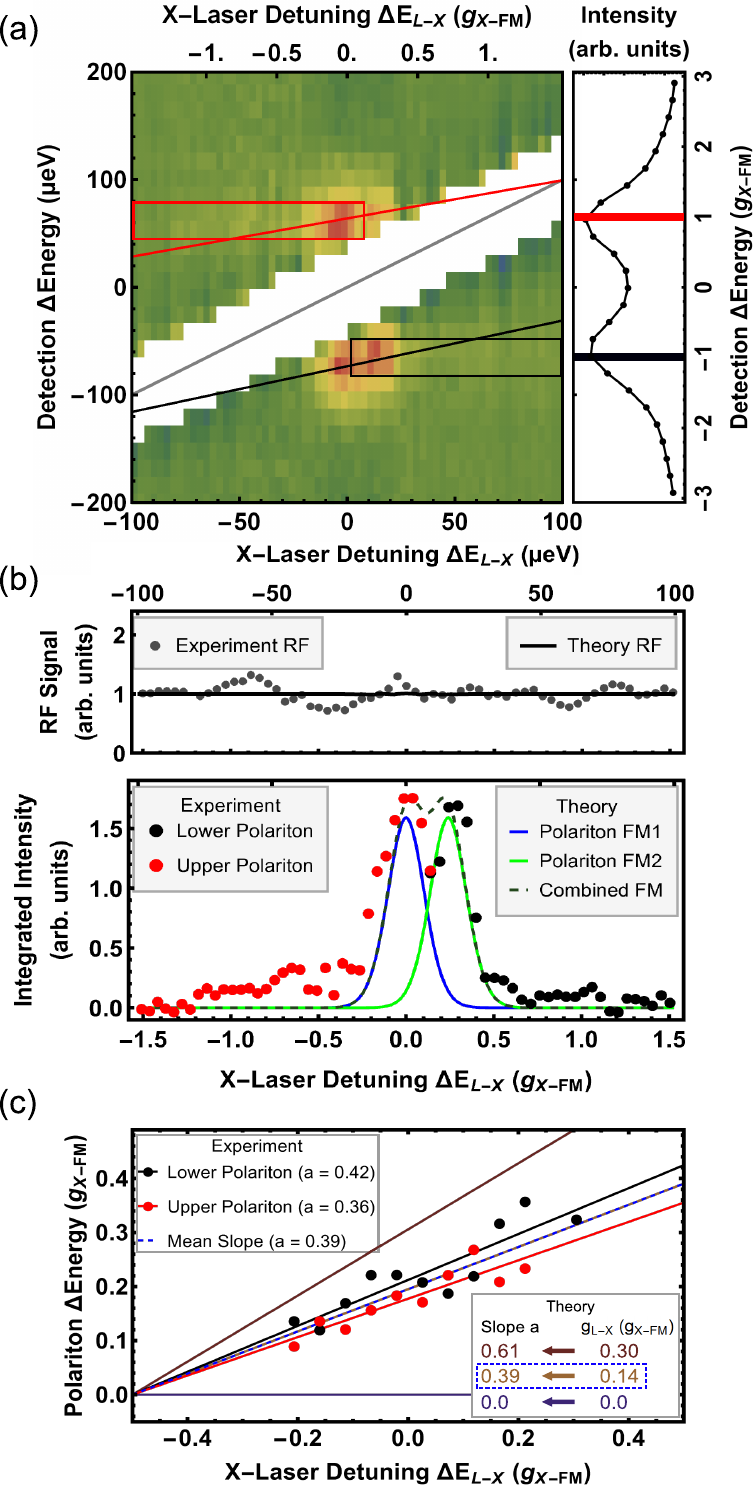}
	\caption{\label{fig:RF_LScan} (a) 2D-map of resonance fluorescence (RF) spectra as a function of the laser-X detuning $\Delta E_\mathrm{L-X}$ at the QD exciton (X) and fundamental mode (FM) resonance ($\Delta E_\mathrm{X-FM} = 0$) using $2$ $nW$ excitation power (measured outside the cryostat). The RF signal is cut from the spectra (white area) for scaling reasons. On the right to the 2D-map a spectrum measured with above-band excitation at $532$ $nm$ is shown for comparison. (b) RF signal and lower (black) as well as upper (red) integrated polariton branch intensities are depicted versus $\Delta E_\mathrm{L-X}$ in the upper and lower panel, respectively. Theoretical curves are shown as an overlay (i.e. no fitting parameters) with solid lines. (c) Upper (red dots) and lower (black dots) polariton energetic shifts and their linear fits (red and black solid lines, respectively) as a function of $\Delta E_\mathrm{L-X}$ together with theoretical predictions for three laser excitation strengths $g_\mathrm{L-X}$ of $0$, $0.14$ and $0.3$ $g_\mathrm{X-FM}$ (solid purple to brown lines). Through comparison between theoretical and experimental slopes $a$ the effective microscopic driving strength $g_\mathrm{L-X}$ can be determined to $0.14$ $g_\mathrm{X-FM}$.}
\end{figure}

Let us first focus on experiments under resonant driving of the cQED system which address the onset of the transitory regime: $g_\mathrm{L-X} \lesssim g_\mathrm{X-FM}$. To characterize the response of the strongly-coupled X-FM system on resonance ($\Delta E_\mathrm{X-FM}=0$) to coherent driving, the laser is scanned across the X-FM resonance and the optical response is recorded (Fig. \ref{fig:RF_LScan}(a)). In the depicted intensity map the RF signal was subtracted for scaling reasons. On the right side of (a) an above-band spectrum is shown for comparison - the two polariton branches marked in red (UP) and black (LP), respectively, can be identified. The intensity of RF signal determined by integration of the detected scattered excitation laser light, is shown in the upper panel of (b). This curve is dominated by the coherent scattering of the excitation laser \cite{Fis2016, Bos2012} and does not show any resonances near $\Delta E_\mathrm{L-X} = 0$ which indicates that the bare X state does not significantly scatter the resonant laser. This can be traced back to the strong coupling of the X to the FM which leads to fast excitation transfer from the bare X state to the X-FM polariton. This interpretation is supported by theoretical calculations (black solid line) where no X related resonances are visible. A X-FM detuning study performed theoretically (not shown here) indicates that the QD has to be detuned as far as $25 \, g_\mathrm{X-FM}$ to restore the RF response at the X transition energy. Strong resonances at the polariton energies for laser tuned to the bare (uncoupled) X state $\Delta E_\mathrm{L-X} = 0$ show that the polaritons are very efficiently pumped through this state. In the lower panel of Fig. \ref{fig:RF_LScan}(b) the integrated intensities (over spectral ranges indicated by colored boxes in Fig. \ref{fig:RF_LScan}(a)) of the polariton branches are depicted as a function of laser detuning. In contrast to cavity-mediated excitation \cite{Gre2015} there is no observable response of UP or LP if the laser is tuned to opposing branches LP and UP, respectively. This clearly demonstrates that under X-driving the system cannot be efficiently excited through the polariton branches, in agreement with the RF signal. To phenomenologically model this observation regarding the excitation efficiency through different channels and its detuning $\Delta E_\mathrm{L-X}$ dependence, an attenuation factor $\mathcal{N}(\gamma_\mathrm{at}, \Delta E_\mathrm{L-X})$ is introduced in the model (Eq. \ref{eq:H_L-X}). Modeling the microscopic origin of this effect is beyond the scope of this work, we attribute it to the orthogonality between laser field and FM wave vectors $\frac{\mathbf{k}}{k}$. On close inspection of the integrated intensities of the polariton branches the central resonance reveals a substructure of $15.6$ $\mu eV$ splitting. This separation matches the FM1-FM2 mode splitting $\delta$FM (measured independently under incoherent pumping) very well. We therefore conclude that both components of the FM - FM1 and FM2 - interact with the bare X independently. Which mode couples to the X strongly is determined by the detuning of the laser with respect to the two modes, whereby the closer FM dominates the X-FM interaction. Interestingly, both FM1 and FM2 resonances feature almost identical intensities, this indicates that preferential polarization axis of the X is equally misaligned to both FM1 and FM2 - i.e. by about $45^\circ$. This interpretation is supported by reports that the X polarization axes of elongated InGaAs QDs are preferentially oriented along the [1-10] and [110] crystal directions \cite{Mus2014}, while the FM components are aligned to the [100] and [010] directions \cite{Rei2007}. Since the theoretical model only considers one FM, there are two independent theoretical curves (solid lines in lower panel in Fig. \ref{fig:RF_LScan}(b)) spaced by the experimentally determined $\delta$FM. The parameters of the attenuation envelope function $\mathcal{N}(\gamma_\mathrm{at}, \Delta E_\mathrm{L-X})$ were chosen for the FWHM of the theoretical resonances to match experimental $\gamma_\mathrm{X} = 0.24$ $g_\mathrm{X-FM}$. Interestingly, we observe driving strength-dependent injection pulling of the X-FM polariton to the resonant laser which is a signature that the resonant laser influences the system significantly. This is the quantum limit of analogous phenomena observed so far for macroscopic systems such as semiconductor lasers \cite{Wie2005}, but not on the level of a single quantum two level system, where no collective effects are present. Observation of this effect of nonlinear dynamics in the regime of cQED requires a joint description thereby bringing the two fields together. Additionally, the observation of injection pulling has also technical applications. It can be utilized to determine the scaling between the driving strength $g_\mathrm{L-X}$ used in calculations (related to the system occupation) and the excitation power $P$ as measured in the experiment. This relation is of great importance for the meaningful comparison, as it links experiment and theory quantitatively, just as illustrated in \ref{fig:RF_LScan}(c), where the relative emission energies of the UP and LP branches are plotted versus laser detuning. The polariton branches follow $\Delta E_\mathrm{L-X}$ with a slope of the linear dependence $a$ of $0.355\pm 0.059$ and $0.425\pm 0.072$ for the UP and LP, respectively. As the slope is expected to change with $g_\mathrm{L-X}$, dependencies for $g_\mathrm{L-X}$ in the range of $0$-$0.3$ $g_\mathrm{X-FM}$ (indicated in the right legend of (c)) were calculated, curves for $g_\mathrm{L-X} = 0$ and $0.3$ $g_\mathrm{X-FM}$ are shown for comparison (purple and brown solid lines). For a theoretical excitation strength of $0.14$ $g_\mathrm{X-FM}$ the resulting slope of $0.39$ is obtained. This matches the experimentally determined mean slope of UP and LP (blue dashed line) of $0.390\pm 0.046$ and therefore enables the identification of the $g_\mathrm{L-X}$ to which the system was subjected to in experiment, when $2$ $nW$ external pump power was applied.


\begin{figure}
	\includegraphics[width=1.0\linewidth]{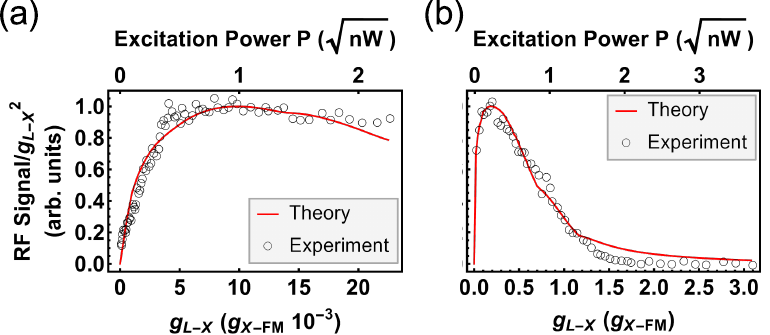}
	\caption{\label{fig:Exp_PScan} Comparison of experimental and theoretical excitation power $P \propto g_\mathrm{L-X}^2$ normalized integrated resonance fluorescence intensities as a function of excitation strength $g_\mathrm{L-FM}$. In the case of (a) for the fundamental mode (FM) detuned with respect to the resonant laser and the quantum dot exciton (X) ($\Delta E_\mathrm{X-FM} = 16$ $g_\mathrm{L-FM}$, $\Delta E_\mathrm{L-X} = 0$) and (b) for the all-resonant case ($\Delta E_\mathrm{X-FM} = \Delta E_\mathrm{L-X} = 0$). The full experimental spectra are presented in appendix \ref{App: Exp Res Power Scans}.}
\end{figure}

Next we investigate the transition between the Jaynes-Cummings and Autler-Towns ladders experimentally by performing excitation power $P$-dependent measurements corresponding to theoretical calculations of Sec. \ref{Sec: Theo}. The resulting RF signals are overlayed with theoretical calculations in Fig. \ref{fig:Exp_PScan} for (a) an off-resonant X-FM system ($\Delta E_\mathrm{X-FM} = 16$ $g_\mathrm{X-FM}$ and $\Delta E_\mathrm{L-X} = 0$) as well as for (b) an all-resonant case ($\Delta E_\mathrm{X-FM}$ = $\Delta E_\mathrm{L-X} = 0$). In experiment and theory the RF signal consists of both, coherent and incoherent response of the system. The full experimental data set is shown explicitly in appendix \ref{App: Exp Res Power Scans}. Both experiment and theory show that when X is off-resonant with respect to the FM (Fig. \ref{fig:Exp_PScan}(a)) the RF signal saturates very fast at a $\sqrt{P}$ of about $0.45$ $\sqrt{nW}$ corresponding to a $g_\mathrm{L-X}$ of $4.5 \cdot 10^{-3}$ $g_\mathrm{X-FM}$. On the other hand, when X is on resonance with FM (Fig. \ref{fig:Exp_PScan}(b)) the experimental and theoretical RF signals exhibit a maximum at $g_\mathrm{L-X}$ equal to $0.22$ $\sqrt{nW}$ are corresponding to $0.18$ $g_\mathrm{X-FM}$, respectively. This maximum has so far not been theoretically described or observed experimentally and can be attributed - as discussed above - to the population of the first manifold of the laser dressed X-FM polariton. It is a robust spectral fingerprint which can be used to identify and pinpoint the intermediate regime even in the case of resolution-limited experimental spectra. For stronger driving the population of higher order rungs of dressed Jaynes-Cummings ladder begin to play a role. Finally, the Mollow triplet dominates the spectrum resulting in a decrease of the optical signal at the laser energy due to increasing contribution of its sidebands. Similarly to the slope of Fig. \ref{fig:RF_LScan}(c) the maximum allows us to determine the relation between the measured experimental excitation power and the theoretical system occupation. Laser-dressed polariton up to the 6th manifold have to be included in order for the theoretical calculations to converge. Since experimental and theoretical curves agree well with each other, this allows us to conclude that also in experiment higher order manifolds of the dressed Jaynes-Cummings ladder contribute to the RF signal.


\section{Conclusion}
\label{Sec: Conclusion}

In conclusion, we describe theoretically and realize experimentally previously unexplored regime of cQED - intermediate between the two well-known limiting cases of incoherently probed Jaynes-Cummings ladder and strongly coherently-driven QD transition - in which laser-dressed polariton states are formed as a result of strong coupling between driving laser, QD X and cavity mode. In this regime none of the interactions can be treated perturbatively and thus description beyond the limiting cases is needed. This is the first realization of this qualitatively distinctive transitory regime, enabled by direct X-driving and the increased cavity losses compared to atomic or superconducting circuit-based QED. Modeling of experimental observations of this transition indicates the first observation of laser-dressed Jaynes-Cummings ladder inherent to direct coupling of the resonant laser to the bare X proven to be a very efficient excitation scheme as opposed to driving the polariton branches directly. This together with observed injection pulling of polariton branches by the external laser paves a way towards realization of nonlinear dynamics phenomena on a single QD level, e.g., cavity-enhanced injection locking. It stresses the importance of the nature (fermionic or bosonic) of the state through which the system is pumped and leads to realization of qualitatively different physical system in the strong driving limit depending whether the driven state undergoes saturation (laser-driven QD) or features unlimited occupation like in the case of cavity-mediated driving. The main and robust spectral fingerprint of the transition is the maximum in the RF signal level with respect to the driving strength proving mutual strong coupling between X, FM and coherent driving field. In the intermediate regime higher splittings within the excitation manifolds (driven by the strength of excitation laser and not limited by the light-matter coupling strength) are combined with the anharmonicity of the Jaynes-Cummings ladder. This has unprecedented implications and shows a feasibility of a continuously tunable anharmonic system as well as switching on the ps time scale between anharmonic (Jaynes-Cummings) and harmonic (Autler-Townes) ladder of states only by using the excitation strength. This can be used in order to coherently prepare and manipulate quantum states. Additionally, we answer the important question of observability of higher order Jaynes-Cummings rungs and the Mollow triplet in strongly-coupled QD-microcavities by defining indispensable prerequisites in terms of excitation scheme and system parameters limiting climbing the Jaynes-Cummings ladder due to its transformation into ladder of double-dressed states under coherent driving.


\begin{acknowledgments}
The research leading to these results has received funding from the German Research Foundation (DFG) via Projects No. Ka2318/4-1 and No. Re2974/3-1, the SFB 787 “Semiconductor Nanophotonics: Materials, Models, Devices”, and from the European Research Council under the European Union's Seventh Framework ERC Grant Agreement No. 615613. A. C. gratefully acknowledges support from SFB 910: "Control of self-organizing nonlinear systems".
\end{acknowledgments}


\appendix

\section{Theoretical Spectra}
\label{App: Theo Spectra}


The system under study consists of a coherently-driven QD coupled strongly to a FM of a microcavity. The electronic structure of the QD is truncated to a two-level system of ground state $ \ket{g}$ and excited state $\ket{e}$ with X transition energy $E_\mathrm{X} $. QD transition and occupation operators are expressed with $\flip{i}{j}=\sigma_{ij}$. The X is driven by a coherent classical laser field with amplitude $g_\mathrm{L-X}$ and energy $ E_L $. Coupling the driving field directly to the electronic subsystem is a distinctive feature of presented modeling in comparison to commonly employed cavity-mediated excitation of the QD. In our case the QD X is coupled to the fundamental mode of the microcavity of $ E_\mathrm{FM} $ energy with coupling strength $g_\mathrm{X-FM}$, but occupation of the cavity ($c^{\dg} c$) is only possible through the electronic states. Respective Hamiltonian $H$ describing this cQED configuration can, in the dipole approximation, be written as:

\begin{align*}
H =& H_0+H_\mathrm{L-X}+H_\mathrm{X-FM}\\
H_0 =& \hbar E_\mathrm{X} \sigma_{ee} + \hbar E_\mathrm{FM} c^\dg c \\
H_\mathrm{L-X} =& g_\mathrm{L-X} \times \mathcal{N}(\gamma_\mathrm{at}, \Delta E_\mathrm{L-X}) \left(e^{i \frac{E_L}{\hbar} t} \sigma_{ge} + e^{-i \frac{E_L}{\hbar} t} \sigma_{eg} \right)\\
H_\mathrm{X-FM} =& g_\mathrm{X-FM} \left( c^\dg \sigma_{ge} + \sigma_{eg} c \right). 
\end{align*}

with $H_0$ corresponding to the electronic and photonic excitations in the system and $H_\mathrm{L-X}$ and $H_\mathrm{X-FM}$ describing coupling of the QD X with the laser and cavity mode, respectively. $\mathcal{N}(\gamma_\mathrm{at}, \Delta E_\mathrm{L-X})$ is an envelope function of a normal distribution of width $\gamma_\mathrm{at}$ which is introduced to reproduce the experimental observation that the excitation of the system through the bare (uncoupled) QD X state is most efficient and it becomes harder with increasing laser detuning $\Delta E_\mathrm{L-X}$. We include this effect phenomenologically as a detuning $\Delta E_\mathrm{L-X}$-dependent attenuation of the laser excitation strength without modeling its origin microscopically, which is beyond the scope of this study. The possible origin of this effect is discussed in Sec. \ref{Sec: Exp} of the main text. The above Hamiltonian is further transformed into the rotating frame of the driving laser frequency and yields the form shown in Eqn. \ref{eq:H} to \ref{eq:H_X-FM}. For the sake of comparison with the experiment, the cavity spectrum is calculated using two-time correlations of the cavity operators:

\begin{align*}
S_\mathrm{FM}(E) =& 
\underset{t\rightarrow\infty}{\mathrm{lim}}
\mathrm{Re}
\left[ 
\int_0^\infty d\tau \ew{ c^\dg(t)c(t+\tau)} e^{-i(\frac{E-E_L}{\hbar})\tau}
\right],
\end{align*}

similarly dipole spectrum can be obtained:

\begin{align*}
S_\mathrm{X}(E) =& 
\underset{t\rightarrow\infty}{\mathrm{lim}}
\mathrm{Re}
\left[ 
\int_0^\infty d\tau \ew{ \sigma_{ge}(t)\sigma_{eg}(t+\tau)} e^{-i(\frac{E-E_L}{\hbar})\tau}
\right]
\end{align*}

This yields the full spectrum centered around the driving laser field energy $ E_L $. For the further discussion of the results it is instrumental to separate the incoherent part from the coherent part which is done by subtracting the coherent part from the full spectrum:


\begin{multline*}
S(E)_\mathrm{incoh} =\\ \underset{t\rightarrow\infty}{\mathrm{lim}} \mathrm{Re}
\left[ 
\int_0^\infty d\tau 
\left(
\ew{ c^\dg(t)c(t+\tau)} 
- \left|\ew{c^\dg(t)}\right|^2
\right)
e^{-i(\frac{E-E_L}{\hbar})\tau}
\right],
\end{multline*}

which allows for an efficient and fast calculation of the incoherent part, as

\begin{align*}
\underset{t,\tau \rightarrow\infty}{\mathrm{lim}}
\left(
\ew{ c^\dg(t)c(t+\tau)} 
- \left|\ew{c^\dg(t)}\right|^2
\right)
=0.
\end{align*}

To calculate the two-time correlations, a numerical approach via a Runge-Kutta integration of the full master von-Neumann equation is chosen with:

\begin{multline*}
\dot \rho = (-i/\hbar)
\left[H_0+H_\mathrm{L-X}+H_\mathrm{X-FM},\rho\right] 
+ \kappa \mathcal{D}[c]\rho \\*
+ \gamma_X \mathcal{D}[\sigma_{ee}-\sigma_{gg}]\rho
+ \Gamma \mathcal{D}[\sigma_{ge}]\rho . 
\end{multline*}

In order to include the dissipative aspects of the investigated system we use the standard Lindblad formulation: $ \mathcal{D}[J]\rho=2J\rho J^\dg - J^\dg J \rho - \rho J^\dg J $. The cavity loss rate is denoted with $\kappa$, pure dephasing of the quantum dot transition with $\gamma_X$ and the X radiative decay rate with $\Gamma$. The dynamics in $ t $ are brought into the steady-state solution with solving $\dot \rho = 0$. Given the steady-state density matrix, the $ \tau $ dynamics are computed via the same master equation, but with an initialization corresponding to the quantum regression theorem:

\begin{align*}
\ew{c^\dg(t) c(t+\tau)}
=&
\mathrm{Tr}\left( \rho(0) c^\dg(t) c(t+\tau) \right)\\
=&
\mathrm{Tr}\left( \rho(0) U(t,0)c^\dg U^\dg(t,0) 
U(t+\tau,0)cU^\dg(t+\tau,0) \right) \\
=&
\mathrm{Tr}\left( U^\dg(t,0) \rho(0) U(t,0) c^\dg 
U(\tau,0)c U^\dg(\tau,0) \right)\\ 
=&
\mathrm{Tr}\left( \rho(t) c^\dg 
U(\tau,0) c U^\dg(\tau,0) \right)\\
=&
\mathrm{Tr}\left(\bar\rho(t+\tau)  c \right). 
\end{align*}

This means that the $\tau$ dynamics are completely governed by the new projected density matrix $\bar\rho(t+\tau)$:

\begin{align*}
\rho(t) c^\dg =&
\sum_{i,j=e,g}
\sum_{m,n=0}^N
c^{im}_{jn}(t) \left|im\right\rangle\left\langle jn\right| \ c^\dg \\
=&
\sum_{i,j=e,g}
\sum_{m,n=0}^N
c^{im}_{jn}(t) \left|im\right\rangle\left\langle jn-1\right|  \sqrt{n} .
\end{align*}

We can define the projected density matrix by relabeling $n^\prime = n-1$

\begin{align*}
\bar\rho(t)=&
\sum_{i,j=e,g}
\sum_{m,n^\prime=0}^N
\bar c^{im}_{jn^\prime}(t) \left|im\right\rangle\left\langle j 
n^\prime \right| 
\qquad
\mathrm{with}\\
\qquad
c^{im}_{jn^\prime}(t) 
=& \sqrt{n+1}c^{im}_{jn+1}(t).
\end{align*}

Given the two-time correlations, the Wiener-Khinchin theorem yields the output spectrum of the driven-cQED system. This approach includes all nonlinear higher-order photon scattering events and is numerically exact.

\section{Analytical Solution For System Eigenvalues}
\label{App: Theo Ana Eigen}

In this section evolution of the system eingenvalues with increasing driving strength is described (cf. Fig. \ref{fig:Schematic_Config} and \ref{fig:Theo_PScan}). Derivation of analytical expressions is possible only when investigated ssytem is greatly simplified and dissipation as well as higher order Jaynes-Cummings rungs are neglected. In the limit of only a single excitation present in the system (either electronic $\ket{e,0}$ or photonic $\ket{g,1}$) the state vector reads: 

\begin{eqnarray*}
	\ket{\psi}
	&=&
	c_g \ket{g,0}
	+
	c_e \ket{e,0}
	+
	c_p \ket{g,1}
\end{eqnarray*} 

with amplitudes: $c_g$ (no excitation in the system), $c_e$ (QD X without photons in the cavity) and $c_p$ (QD in the ground state $\ket{g}$ and 1 photon in the cavity).

In the rotating frame of the laser and in the corresponding basis the Hamiltonian can be written as:

\begin{eqnarray*}
	H
	&=&
	\begin{pmatrix}
		0 & 0 & 2 g_\mathrm{L-X} \\
		0 & E_\mathrm{FM} & g_\mathrm{X-FM} \\
		2 g_\mathrm{L-X} & g_\mathrm{X-FM} & E_X 
	\end{pmatrix}.
\end{eqnarray*}

By diagonalization of this 3x3 matrix we yield the following eigenenergies:

\begin{eqnarray*}
	\lambda_1
	&=& \frac{1}{3}
	\left(
	b + 2\sqrt{p} \ \cos\left( \frac{E_L}{3} \right)
	\right) \\
	\lambda_2
	&=& \frac{1}{3}
	\left(
	b + 2\sqrt{p} \ \cos\left( \frac{E_L+2\pi}{3} \right)
	\right) \\
	\lambda_3
	&=& \frac{1}{3}
	\left(
	b + 2\sqrt{p} \ \cos\left( \frac{E_L-2\pi}{3} \right)
	\right), 
\end{eqnarray*}

with the following definitions introduced:

\begin{eqnarray*}
	b
	&=& E_\mathrm{FM} + E_X \\
	c &=& E_\mathrm{FM} E_X - g_\mathrm{X-FM}^2 - 4 g_\mathrm{L-X}^2 \\
	d &=& E_\mathrm{FM} g_\mathrm{X-FM}^2.
\end{eqnarray*}

Taking the general solution for real symmetric 3x3 matrix we yield:

\begin{eqnarray*}
	E_L
	&=& \mathrm{arccos}\left( \frac{q}{2\sqrt{p^3}} \right)
\end{eqnarray*} 

with

\begin{eqnarray*}
	q 
	&=&
	2b^3 - 9bc -27d \\
	p &=& b^2 - 3c.
\end{eqnarray*}

The numerical solutions correspond to the analytical solutions. In the case when laser, FM of the cavity and QD X are all in resonance $E_L = E_X = E_\mathrm{FM}$ the evolution of the eigenenergies $\lambda_1$ and $\lambda_2$ corresponding to the upper and lower polariton UP and LP, respectively, reduces to:

\begin{eqnarray}
\lambda_\mathrm{UP}
&=& +\sqrt{g_\mathrm{X-FM}+ 4 g_\mathrm{L-X}^2} \label{eq:UP_Eigenvalue} \\
\lambda_\mathrm{LP}
&=& -\sqrt{g_\mathrm{X-FM}+ 4 g_\mathrm{L-X}^2}. \label{eq:LP_Eigenvalue}
\end{eqnarray}

\section{Emission Spectra as a Function of Resonant Excitation Power}
\label{App: Exp Res Power Scans}

\begin{figure}
	\includegraphics[width=1.0\linewidth]{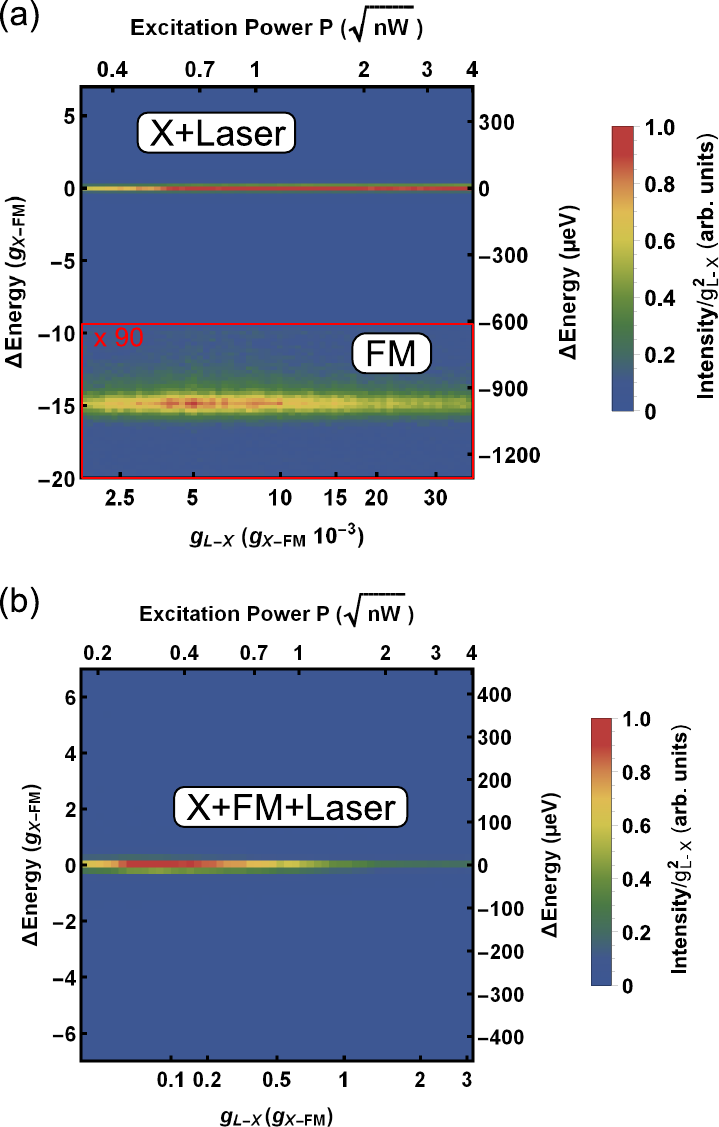}
	\caption{\label{fig:Suppl_ExpPower} Experimental resonance fluorescence (RF) spectra under resonant excitation of quantum dot exciton (X) as a function of the square root of excitation power $\sqrt{P}$ ($\propto$ excitation strength $g_\mathrm{L-X}$). In panel (a) the fundamental mode (FM) is detuned by about $-15$ $g_\mathrm{X-FM}$ from the X (and excitation laser), in (b) X, FM and laser are on resonance. (a) and (b) correspond to the integrated intensities shown in Fig. \ref{fig:Exp_PScan}(a) and (b), respectively. This data was also employed to infer the scaling between $\sqrt{P}$ and $g_\mathrm{L-X}$.} 
\end{figure}

In Fig. \ref{fig:Exp_PScan} the experimental integrated intensities for an off- and an on-resonant QD X-FM case are compared to theory. The corresponding power-dependent experimental spectra from which these integrated intensities are extracted are presented in Fig. \ref{fig:Suppl_ExpPower}. The detuned FM case of (a) and the all-resonant case of (b) correspond to Fig. \ref{fig:Exp_PScan}(a) and (b), respectively. Panel (b) also corresponds to the theoretical graph shown in Fig. \ref{fig:Theo_PScan}. The difference between the two is that the theory graph shows only the incoherent contribution while panel (b) includes both coherent and incoherent contributions (not separable in experiment) together with residual laser stray-light. To eliminate the contribution of the excitation stray-light experimental emission intensities are normalized by the excitation power $P \propto g_\mathrm{L-X}^2$ since it is expected to scale linearly with $P$. In panel (a) the FM is detuned from X and the laser by about $-15$ $g_\mathrm{X-FM}$. The normalized intensities around the FM are magnified by a factor of $90$ to show that long-range off-resonant X-FM coupling is present as reported in \cite{Ate2009a, Maj2011}. The Mollow triplet predicted by theory can not be resolved experimentally due to large QD exciton emission FWHM of about $15$ $\mu eV$ and high intensity of the resonant scattering (and stray-light) at $\Delta E_\mathrm{L-X} = 0$ - which is at least 4 orders of magnitude higher than non-resonant features.

\bibliography{Bibliography}

\end{document}